\documentstyle{article}

\font\tenrm=cmr10
\font\tenit=cmti10
\font\elevenbf=cmbx10 scaled\magstep 1
\font\elevenrm=cmr10 scaled\magstep 1
\font\elevenit=cmti10 scaled\magstep 1
\font\ninebf=cmbx9
\font\ninerm=cmr9

\begin{document}
\begin{flushright}
hepth/9306095
\end{flushright}

\begin{center}
\vglue 0.6cm
 {\elevenbf  \vglue 10pt
               SUPERSYMMETRY AND BLACK HOLES \footnote{
\ninerm\baselineskip=11pt Invited talk given at SUSY 93, Boston,
April
1993.} \\}
\vglue 5pt

\vglue 1.0cm
{\tenrm RENATA KALLOSH  \\}
\baselineskip=13pt
{\tenit Physics Department, Stanford University, \\}
\baselineskip=12pt
{\tenit Stanford, CA 94305\\}
\end{center}

\vglue 0.3cm
{\rightskip=3pc
 \leftskip=3pc
 \tenrm\baselineskip=12pt
\vglue 0.8cm
\noindent We consider  static black
holes, which are bosonic solutions of supersymmetric theories. We
will show that
supersymmetry provides a natural framework for a discussion of
various
properties of such
 static black holes. The most fundamental property of simple global
supersymmetry,
non-negativeness of the energy, is generalized in the black hole
theory to
cosmic censorship. The SUSY classification of  static black holes
will be presented in terms of central charges of supersymmetry
algebra.
The mass, the temperature and the
entropy of these black holes are simple functions of supersymmetry
charges.
The extreme black holes have zero temperature and some unbroken
supersymmetries.

\vglue 0.8 cm

Simple $N=1$ global supersymmetry algebra relates supercharges to the
Hamiltonian $H=P^0$.
\begin{equation}
\{Q_{\alpha}, \bar Q_{\dot \beta}\}= (\sigma_\mu)_{\alpha \dot
\beta}{} P^\mu~.
\end{equation}
This implies that the spectrum of the Hamiltonian is semipositive
definite
 since $\bar Q$ is Hermitean adjoint of $Q$ and we have
\begin{equation}
H=P^0 =
 (\bar Q_1 Q_1 + Q_1\bar Q_1+\bar Q_2 Q_2 + Q_2\bar Q_2)\equiv |Q|^2
\geq 0~.
\end{equation}
Local supersymmetry does not require positivity of energy, in
general. However,
if the curved space is asymptotically flat, states at infinity may be
described
by the multiplets of global supersymmetry. In particular, a
generalization of the
 positiveness of the energy has been found  for static black
holes, considered as bosonic solutions of supersymmetric theories
in $^{1, 2}$. Axion-dilaton
$U(1)^2$ black holes of a mass $M$ in four-dimensional theory are
associated,
as shown in ref.
 $^{2}$ with  extended $N=4$ supersymmetry algebra with central
charges.

\begin{eqnarray}
\{Q^{I}_{\alpha},Q^{* J}_{\dot{\beta}}\}&=&
(\sigma_{\mu})_{\alpha\dot{\beta}}P^{\mu}\delta^{IJ}\ , \\
\{Q^{I}_{\alpha},Q^{J}_{\beta}\}&=&\epsilon_{\alpha\beta}Z^{IJ}\ ,
\hspace{2cm}
Z^{IJ}=-V^{IJ}+iU^{IJ}\\
\{Q^{*I}_{\dot{\alpha}},Q^{J}_{\dot{\beta}}\}
&=&\epsilon_{\dot{\alpha}\dot{\beta}}\bar Z^{IJ}\,   \ .
\end{eqnarray}
In the rest frame in a special basis this algebra describes a system
of states,
characterized by the mass and two complex central charges $z_1$ and
$z_2$.
 \begin{eqnarray}
 \{q_{(1)} , q_{(1)}^{*}\}&=& 2|q_{(1)}|^2 = M - |z_1|  \geq
0 \ , \nonumber\\
 \{q_{(2)} , q_{(2)}^{*}\}&=& 2|q_{(2)}|^2 = M - |z_2|  \geq
0 \ ,\nonumber\\
  \{q_{(3)} , q_{(3)}^{*}\}&=&2|q_{(3)}|^2 = M + |z_1|  \geq 0 \ ,
\nonumber\\
\{q_{(4)} , q_{(4)}^{*}\}&=&2|q_{(4)}|^2 = M + |z_2|  \geq 0 \ .\\
\label{z}
\end{eqnarray}
The central charges are functions of electric $Q$ and magnetic $P$
charges of the
vector field $F_{\mu\nu }^1$ and axial-vector field
$F_{\mu\nu}^2$.
\begin{eqnarray}
z_{1}&=&\sqrt{2}(\Gamma_{1}+i\Gamma_{2}) \ , \\
z_{2}&=&\sqrt{2}(\Gamma_{1}-i\Gamma_{2}) \  ,
\end{eqnarray}
where $\Gamma=\frac{1}{2}(Q+iP)$. The energy ( the mass) of the
charged black hole is not only non-negative but according to
supersymmetry algebra (\ref{z}) it has to be greater or equal to
central charges.
\begin{eqnarray}
M \geq |z_1| \ , \qquad M \geq |z_1| \ .
\label{lim}\end{eqnarray}
It is quite surprising that for all known to us static black holes in
supersymmetric theories the limits (\ref{lim}), required by
supersymmetry,
coincide exactly with the limits on the parameters of the black holes
required by
the cosmic censorship conjecture. The necessary condition is the
possibility of
embedding the bosonic action into a supersymmetric action. For
example, for the
black holes which are obtained from the action where dilaton
interacts
with vector fields as $e^{2a\phi} F_{\mu\nu}^2$ and $a$ is an
arbitrary parameter,
we do not have any information of that kind except when $a=1$
(or $\sqrt 3$) and supersymmetric embedding is known.

The most important parameters describing the
black holes are its mass and charges. In addition, one usually
calculates the
Hawking temperature and the Bekenstein-Hawking entropy of a black
hole, which are usually some functions of the mass and charges. We
have found
that  all these  parameters of a black hole can be expressed in terms
of the supercharges, describing this system. In particular, the mass
is given by one
half of the
sum over all four supercharges $ |q_{(1)}|, \;|q_{(2)}|, \;|q_{(3)}|,
\;|q_{(4)}|$.
\begin{equation}
M =\frac{1}{2}\sum_{i=1}^{i=4}|q_{(i)}| ^2\ .
\label{mass}\end{equation}

The product of the temperature and entropy is proportional  to the
product of all four
supercharges.
\begin{equation}
S\;T = { 4 \prod_{i=1}^{i=4}|q_{(i)}| \over
\sum_{j=1}^{j=4}|q_{(j)}|^2}
\end{equation}
The temperature is also a function of supercharges.

\begin{eqnarray}
T=\frac{1}{\pi\sum_{j=1}^{j=4}|q_{(j)}|^2}\;
\frac{\prod_{i=1}^{i=4}|q_{(i)}|}{(|q_{(1)}q_{(3)}|+|q_{(2)}q_{(4)}|)^
{2}}~.
\label{T}\end{eqnarray}

The Hawking temperature of the black hole, according to  this
equation,
vanishes  if one of the eigenvalues of the 4 supercharges is zero.
However, if two supercharges approach zero eigenvalues, the limiting
temperature
depends on the order in which this limit is taken. If first we assume
that for
example $|q_{(1)}| \rightarrow |q_{(2)}|$ and $|q_{(1)}| \rightarrow
|q_{(2)}|$, the
limiting temperature equals $T=\frac{1}{8\pi M}$. This correspond to
the limit
from $U(1)^2$ black hole with two vector fields to  $U(1)$ black hole
with one
vector field. If however one first take the limit $|q_{(1)}|
\rightarrow 0$ at fixed
$ |q_{(2)}|$, the temperature goes to zero and remains zero when
the limit $|q_{(2)}| \rightarrow 0$ is taken afterwards.\\

The  entropy can be represented as   a  function of supercharges:
\begin{eqnarray}
S&=&4\pi(|q_{(1)}q_{(3)}|+|q_{(2)}q_{(4)}|)^{2}~.
\label{S}\end{eqnarray}

As it follows from this equation, the entropy vanishes when
$|q_{(1)}| = |q_{(2)}| = 0$. The zero eigenvalue of the supercharge
indicates the
existence of unbroken supersymmetries.

The fact that the mass, the temperature and the entropy of static
black holes can be
completely described in terms of the eigenvalues of the supercharges
is quite
remarkable! This is a property of all static black holes and not just
of extreme
black holes which have some unbroken
 supersymmetries.

\vskip 1 cm

Supersymmetry provides a simple classification of four-dimensional
static black holes, starting with classical Schwarzschild and
Reissner-Nordstr\"om black holes up to  axion-dilaton black holes.

Our conventions and  are those of
$^{2}$.  We
will use the complex scalar $\lambda = a+ie^{-2\phi}$,
where $a$ is the axion field and $\phi$
is the dilaton field. We also have two  $U(1)$ vector
fields $A_{\mu}^{(n)}$, $n=1,2$.

We find it convenient to define the
$SL(2,R)$-duals\footnote{The spacetime duals are
${}^{\star}F^{(n)\mu\nu}=\frac{1}{2\sqrt{-g}}
\epsilon^{\mu\nu\rho\sigma}F_{\rho\sigma}$, with
$\epsilon^{0123}=\epsilon_{0123}=+i $  .} to the
fields $F_{\mu\nu}^{(n)}= \partial_{\mu} A_{\nu}^{(n)}-
\partial_{\nu}
A_{\mu}^{(n)}$\ ,
\begin{equation}
\tilde{F}^{(n)}=e^{-2\phi}\,{}^{\star}F^{(n)}-iaF^{(n)}\; ,
\end{equation}
in terms of which the bosonic part of  $N=4$ supergravity action
reads
\begin{equation}\label{eq:action1}
S=
\frac{1}{16\pi}
\int d^{4}x\sqrt{-g}\biggl
\{-R+\frac{1}{2}\frac{\partial_{\mu}\lambda
\partial^{\mu}\overline{\lambda}}{({\mbox{Im}} \; \lambda)^{2}}
-\sum_{n=1}^{2}F^{(n)}_{\mu\nu}{}^{\star}
\tilde{F}^{(n)\mu\nu} \biggr \}\; .
\end{equation}
In terms of the component fields, we have
\begin{eqnarray}\label{eq:action2}
S & = &
\frac{1}{16\pi}
\int d^{4}x\sqrt{-g}\biggl \{-R+2(\partial\phi)^{2}
+\frac{1}{2}e^{4\phi}(\partial a)^{2}-
\nonumber \\ &&
-e^{-2\phi}\sum_{n=1}^{2}(F^{(n)})^{2}
+ia\sum_{n=1}^{2}F^{(n)}{}^{\star}F^{(n)}\biggr \}\; .
\end{eqnarray}

The advantage of using $\tilde{F}^{(n)}$ is that the
equations of motion imply the local existence of $N$
real vector potentials $\tilde{A}^{(n)}$ such that
\begin{equation}
\tilde{F}^{(n)}=i\, d\tilde{A}^{(n)}\; .
\end{equation}
The analogous equation
$F^{(n)}=dA^{(n)}$ is not a consequence of  equations
of motion but a consequence of the Bianchi identity.
If the time-like components $A_{t}^{(n)}$
play the role of electrostatic
potentials, then the $\tilde{A}^{(n)}_{t}$  will play the role of
magnetostatic potentials. The $SL(2,R)$ duality transformations
consist in the mixing of $A^{(n)}$ with $\tilde{A}^{(n)}$ and
of equations of motion with Bianchi identities,
as in the Einstein-Maxwell case.

Here we present two different kinds of static
solutions to the equations of motion of the action
(\ref{eq:action1}), (\ref{eq:action2}): spherical
black-hole solutions and multi-black-hole
solutions, both with nontrivial axion, dilaton and
$U(1)$ fields. All the previously
known solutions of these kinds (Schwarzschild, (multi-)
Reissner-Nordstr\"{o}m, the  electric and  magnetic
dilaton black holes,
 the electric-magnetic axion-dilaton black holes of refs.$^{3, 2}$
are particular cases of
them.

The static spherically symmetric black-hole solutions are$^2$
\begin{eqnarray}
ds^{2}                 & = &
e^{2U}dt^{2}-e^{-2U}dr^{2}-R^{2}d\Omega^{2}\; ,
\nonumber \\
\nonumber \\
\lambda(r)             & = &
\frac{\lambda_{0}r+\overline{\lambda}_{0}\Upsilon}{r
+\Upsilon}\; ,
\nonumber \\
\nonumber \\
A_{t}^{(n)}(r)         & = &
e^{\phi_{0}}R^{-2}[\Gamma^{(n)}(r+\Upsilon)+c.c]\; ,
\nonumber \\
\nonumber \\
\tilde{A}_{t}^{(n)}(r) & = &
-e^{\phi_{0}}R^{-2}[\Gamma^{(n)}(\lambda_{0}r+
\overline{\lambda}_{0} \Upsilon)+c.c]\; ,
\label{solution}\end{eqnarray}
where
\begin{eqnarray}
e^{2U}(r)  & = & R^{-2}(r-r_{+})(r-r_{-})\;,  \qquad
r_{\pm}    = M\pm r_{0}\; ,
\nonumber \\
R^{2}(r)   & = &  r^{2}-|\Upsilon|^{2}\; ,
\hspace{2,8 cm}r_{0}^{2}=M^{2}+|\Upsilon|^{2}
-4\sum_{n=1}^{N} |\Gamma^{(n)}|^{2}\; .
\end{eqnarray}

We define the parameters of our solutions
in terms of the asymptotic behavior ($r\rightarrow
\infty$) of   different complex fields
\begin{eqnarray}
g_{tt} & \sim  & 1-\frac{2M}{r}\; ,
\hspace{3cm}
\lambda \sim \lambda_{0}-ie^{-2\phi_{0}}
\frac{2\Upsilon}{r}\; , \nonumber \\
F_{tr}  & \sim
&\frac{e^{+\phi_{0}}Q}{r^{2}}\; . \hspace{3.3cm}
{}^{\star}F_{tr}  \sim \frac{i
e^{+\phi_{0}}P}{r^{2}}\; .
\end{eqnarray}

The real axion ($\Delta$), dilaton ($\Sigma$), electric
($Q$) and magnetic ($P$) charges, and the
asymptotic values of the
axion ($a_{0}$) and dilaton ($\phi_{0}$) are
\begin{equation}
\Upsilon  = \Sigma-i\Delta,
\hspace{1cm}
\Gamma   = \frac{1}{2}(Q+iP),
\hspace{1cm}
\lambda_{0}  =  a_{0}+ie^{-2\phi_{0}}.
\end{equation}

In every black hole described by our solutions the charge of
the complex scalar  $\Upsilon$ is related to the electromagnetic
charges by
\begin{equation}
\Upsilon=- \frac{2
}{M}\, (\overline{\Gamma}_{n})^{2}\; .
\end{equation}

The singularity is hidden under the horizon if $r_{0}^{2}>
0$, and it is hidden or coincides with it (but still is
invisible for an external observer) if $r_{0}=0$.
The  conditions
$r_{0}^{2} \geq 0$
 and $M\geq |\Upsilon|$ can be related to supersymmetry bounds
$^{2}$. All solutions given above have the entropy
\begin{equation}
S= \pi (r_{+}^2 - |\Upsilon|^{2}) \ .
\end{equation}
When all  supersymmetric bounds are saturated, i.e.
$r_{+}=M=|\Upsilon|$,
the objects described by this solution have
 zero area of the horizon and vanishing entropy.  In this sense, such
black holes ({\it holons})  behave as elementary particles.

The electric, magnetic, dilaton and axion charges can be expressed in
terms of
the central  charges $z_{1},z_{2}$  and the mass $M$:
\begin{eqnarray} \Gamma_{1}&=&\frac{1}{2\sqrt{2}}(z_{1}+z_{2})\ , \\
\Gamma_{2}&=&\frac{1}{2\sqrt{2}i}(z_{1}-z_{2})\ , \\
\bar \Upsilon&=&\Sigma+i\Delta=-\frac{z_{1}z_{2}}{M}~.
\\ \end{eqnarray}

Thus our axion-dilaton black holes have the following ``hair": the
mass $M$, two
complex central charges $z_1, z_2$ (through which electric, magnetic,
 dilaton
and axion charges are expressed) and asymptotic value at infinity of
the dilaton
and of the axion field $\lambda_0 = a_0 + e^{-2\phi_0}$.

The  temperature of a whole family of
axion-dilaton black holes is
\begin{equation}
T =   {1 \over 2\pi M} \quad
 {\sqrt{(M^2 - |z_1|^2) \; (M^2 - |z_2|^2)}\over \left[ \sqrt{M^2 -
|z_1|^2 } +  \sqrt{ M^2 - |z_2|^2}\right ]^2}\ .
\label{tem}\end{equation}

The entropy, which equals  one quarter of the area of the horizon, is

\begin{equation}
 S =  \pi \left[ \sqrt{M^2 - |z_1|^2 } +  \sqrt{ M^2 - |z_2|^2}\right
]^2 ~.
\label{en}\end{equation}

The parameter $r_0$, which vanishes when the black hole becomes
extremal, is given by the product of supersymmetry bounds:

\begin{equation}
r_{0}^2 =\frac{(M^2 -
|z_1|^2)\,
(M^2 - |z_2|^2)}{M^2} \  .
 \end{equation}

\vskip 0.8 cm
SUSY CLASSIFICATION OF STATIC BLACK HOLES will be given in terms of
 central charges and asymptotic values of dilaton-axion field at
infinity.

\vskip 0.5 cm
\noindent {\bf Schwarzschild}  solution
corresponds
to $z_1 = z_2 =\lambda_0 =i$.  Thus we get $r_0=M, \;r_+=2M,
\;r_-=0$,
the vector and dilaton-axion fields are absent in this solution. The
temperature
and entropy
are obtained from eqs. (\ref{tem}), (\ref{en}).
 \begin{eqnarray}
T_{Sch} &= & {1 \over 8\pi M} ~,\nonumber\\
\nonumber\\
S_{Sch} &=& 4 \pi M^2\ .
\end{eqnarray}
{\bf Reissner-Nordstr\"om} solution is given by $z_1=\lambda_0 =i, \;
z_2 = q$.
We have $r_0=  \sqrt {M^2 - q^2} , \;r_+=2M, \;r_-=0$,
the vector and dilaton-axion fields are absent in this solution.
Substitution
of these values of central charges into eqs.  (\ref{tem}), (\ref{en})
gives
\begin{eqnarray}
T_{RN} &= & {1 \over 2\pi } \;{ \sqrt {M^2 - q^2}  \over \left (M +
 \sqrt {M^2 - q^2} \right )^2}~,\nonumber\\
\nonumber\\
S_{RN} &=& \pi  \left (M +
 \sqrt {M^2 - q^2} \right )^2\ .
\end{eqnarray}
{\bf Extreme Reissner-Nordstr\"om} black holes with $M=q$ have zero
temperature
and  non-zero entropy, $N=1$ supersymmetry being unbroken$^1$.

\noindent   $U(1)$ {\bf Charged  axion-dilaton   black
holes} $^{3}$ have $|z_1| = |z_2| = |Q/ \sqrt 2|$. The corresponding
expressions for
the temperature and the
entropy are
\begin{eqnarray}
T_{U(1)} &= & {1 \over 2\pi M} \;{ M^2 - |Q / \sqrt 2|^2 \over 4 (M^2
-
|Q/ \sqrt 2|^2
)}= {1 \over 8\pi M} \ ,\nonumber\\
\nonumber\\
S_{U(1)} &=& 4 \pi  \left (M^2 - |Q / \sqrt 2|^2 \right )^2\ .
\end{eqnarray}
{\bf In the extreme limit} $M = |Q / \sqrt 2|$ $N=2$ supersymmetry
is restored. The temperature
remains equal to the
temperature of the  Schwarzschild black hole (however, see the
discussion
above where it is explained that this temperature takes different
values if the limit
to extreme is taken starting from $U(1)\times U(1)$ configuration)
and the entropy vanishes.

Finally consider an {\bf $U(1)\times U(1)$ four-dimensional
axion-dilaton black holes}, with one vector and one axial vector
field.
The central charges in terms of
electromagnetic charges are given by
\begin{eqnarray}\label{central}
|z_1|^2 &=& \frac{1 }{ 2}(P_1^2 + Q_1^2 + P_2^2 + Q_2^2) - E  \
,\nonumber\\
\nonumber\\
|z_2|^2 &=& \frac{1 }{ 2}(P_1^2 + Q_1^2 + P_2^2 + Q_2^2)+E \ ,
\end{eqnarray}
where
\begin{equation}
E\equiv Q_1 P_2 - Q_2 P_1\ .
\end{equation}
$Q_1, P_1$ and $Q_2, P_2$ are electric and magnetic charges of
a vector and axial-vector fields, respectively.
The entropy and temperature of these solutions are
given in eqs. (\ref{tem}),  (\ref{en}).

{\bf Extreme charged  $U(1)\times U(1) $ axion-dilaton black holes}
have a
restored
$N=1$ supersymmetry.
 Extreme solution $M=|z_1|$ or $M=|z_1|$, $|z_1| \neq |z_2|$   has
zero temperature
and non-zero entropy,
\begin{eqnarray}
T&=&0 \ ,
\nonumber\\
\nonumber\\
S&= & =   \pi \;(|z_1|^2
- |z_2|^2 ) = \pi (Q_1 P_2 - Q_2 P_1)~.
\end{eqnarray}

There are different  ways to reach the restoration of $N=2$
supersymmetry. One possibility is to consider
the next step $|z_1| \rightarrow  |z_2|$.
In this particular limit
\begin{eqnarray}
T&=&0\ ,
\nonumber\\
S&= &0\ .
\end
{eqnarray}
and the system seems to approach the ground state with zero
temperature and
entropy.

The analogous situation was discussed at length
 and illustrated in figures in ref. {2}, where we plotted both
temperature and
entropy either as the functions of central charges at fixed mass or
as the
function of mass at a given central charges.
 All figures presented there for the dilaton solution apply to the
more
general axion-dilaton solutions under the condition that   instead
of the
central charges one should use the moduli of complex central charges,
since only
in absence of axion are the central charges real.

Our second
kind of solutions describe  axion-dilaton extreme multi-black-hole
solutions$^2$. {\bf  Multi-black-holes have unbroken
supersymmetries}.
 The fields are
\begin{eqnarray}
ds^{2}           & = & e^{2U}dt^{2}-e^{-2U}d\vec{x}^{2}\;
, \hspace{1cm}
                                e^{-2U}(\vec{x})=2 \;{\mbox{Im}}\;
                             ({\cal H}_{1}(\vec{x})\;
                                 \overline{{\cal
                                H}}_{2}(\vec{x}))\; ,
                                    \nonumber \\
\nonumber \\
\lambda(\vec{x}) & = & \frac{{\cal H}_{1}(\vec{x})}{{\cal
                         H}_{2}(\vec{x})}\; ,
                        \nonumber \\
\nonumber \\
A_{t}^{(n)}(\vec{x})       & = & e^{2U}(k^{(n)}{\cal
                                H}_{2}(\vec{x})+c.c)\; ,
                        \nonumber \\
\nonumber \\
\tilde{A}^{(n)}_{t}(\vec{x}) & = & -e^{2U}(k^{(n)}{\cal
                                 H}_{1}(\vec{x})+c.c)\; ,
\end{eqnarray}
where ${\cal H}_{1}(\vec{x}), {\cal H}_{2}(\vec{x})$
are two complex harmonic functions
\begin{eqnarray}
{\cal H}_{1}(\vec{x}) & = & \frac{e^{\phi_{0}}}{\sqrt{2}}
                            \{\lambda_{0}+\sum_{i=1}^{I}
                            \frac{\lambda_{0}M_{i}+
                            \overline{\lambda}_{0}
                            \Upsilon_{i}}{|\vec{x}
                            -\vec{x}_{i}|}\}\; ,
                                      \nonumber \\
\nonumber \\
{\cal H}_{2}(\vec{x}) & = & \frac{e^{\phi_{0}}}{\sqrt{2}}
                            \{1+\sum_{i=1}^{I}\frac{M_{i}+
                            \Upsilon_{i}}{|\vec{x}
                            -\vec{x}_{i}|}\}\; .
\end{eqnarray}

The horizon of the $i$-th (extreme) black hole is at
$\vec{x}_{i}$ (in these
isotropic coordinates the horizons look like single
points), and has mass $M_{i}$, electromagnetic
charge $\Gamma_{i}$, etc. , as can be seen by using the
definitions in the Appendix
in the limit $|\vec{x}-\vec{x}_{i}|\rightarrow\infty$.
Charges without
label are total charges. The constants $k^{(n)}$ are
\begin{equation}
k^{(n)}=
-\sqrt{2}
\biggl (\frac{\Gamma^{(n)}M+\overline{\Gamma^{(n)}
\Upsilon}}{M^{2}-|\Upsilon|^{2}}\biggr )\; .
\end{equation}
The consistency of the solution requires for every $i$
\begin{equation}
k^{(n)}_{i}=k^{(n)}\; ,
\hspace{1cm}
Arg(\Upsilon_{i})=Arg(\Upsilon)\; .
\end{equation}
Finally, for each $i$ and also for the total charges, the
supersymmetric Bogomolny bound is saturated:
\begin{equation}
M^{2}+|\Upsilon|^{2}-
4|\Gamma^{(n)}|^{2}=0\; .
\end{equation}

For a single $U(1)$ vector field
 the extreme solution simplifies to ($k\equiv k^{1}$, $\Gamma \equiv
\Gamma^{1}$)
\begin{equation}
M^{2}=|\Upsilon|^{2}\ , \quad k
 =  -\frac{ \sqrt{2}  \;\Gamma}{ M}
\; .\end{equation}
As a consequence
of all the identities obeyed by the charges, it is
possible to derive the following expression of equilibrium
of forces between two extreme black holes$^2$:
\begin{equation}\label{eq:equiforce}
M_{1}M_{2}+\Sigma_{1}\Sigma_{2}+\Delta_{1}\Delta_{2}=
Q_{1}Q_{2}+
P_{1}P_{2}\; .
\end{equation}

 \vskip 1 cm

If supersymmetry will be discovered in the coming experiments, it
will be
much more natural to study non-perturbative aspects of quantum
gravity in the
context of supersymmetric theories. At the present time
supersymmetric approach
to quantum gravity
is motivated mainly by its internal consistency and existence of
natural
structures describing such interesting objects as black holes. An
important advantage of this approach to the theory of black holes is
the possibility to obtain results which in certain cases do not
acquire any quantum corrections$^2$.

In this paper we have discussed many properties of  extreme and
non-extreme
black holes in terms of  supercharges.
Our main conclusion is that in supersymmetric theories evaporation of
black holes
is related to restoration of supersymmetry.

\vskip 1 cm

This work was
supported by the NSF grant PHY-8612280 and in part by
Stanford University.

 \vglue 1 cm

\noindent {\bf References}

\begin{enumerate}
\item \label{GH} G.W. Gibbons and C.M. Hull, {\elevenit Phys. Lett.}
{\elevenbf109B},    (1982) 190.
\item \label{US}
        R. Kallosh, A. Linde, T. Ort\'{\i}n, A. Peet and
        A. Van Proeyen, {\elevenit Phys. Rev.} {\elevenbf D46}, 5278
(1992);\\
  R. Kallosh  and T. Ort\'{\i}n,   {\elevenit Phys. Rev.} {\elevenbf
D48}, 713 (1993);\\
R. Kallosh  and T. Ort\'{\i}n, {\elevenit  Killing Spinor
Identities}, preprint SU-ITP-93-16
(hepth/9306085).
\item \label{G}
        G. W. Gibbons, {\elevenit Nucl. Phys.} {\elevenbf B207}
(1982) 337;\\
        G. W. Gibbons and K. Maeda, {\elevenit Nucl. Phys.}
{\elevenbf B298}  (1988)        741;\\
        D. Garfinkle, G. Horowitz and A. Strominger,
        {\elevenit Phys. Rev.} {\elevenbf D43}  (1991) 3140;\\
 A. Shapere, S. Trivedi and F. Wilczek,
 {\elevenit Mod. Phys. Lett.}  {\elevenbf A6}  (1991) 2677.
\end{enumerate}

\end{document}

{\ninebf Footnotes} should be typeset in 9 point roman
at the bottom of the page where it is cited.}

\end{document}